\documentclass[conference]{IEEEtran}
\IEEEoverridecommandlockouts
\usepackage{cite}
\usepackage{amsmath,amssymb,amsfonts}
\usepackage{algorithmic}
\usepackage{graphicx}
\usepackage{textcomp}
\usepackage{xcolor}

\usepackage{xcolor}
\usepackage{amsmath}
\usepackage{threeparttable}
\usepackage{siunitx}
\usepackage{glossaries}
\usepackage{amssymb}
 \usepackage{booktabs} 
 \usepackage{multirow}
\usepackage{listings}
\usepackage{courier}
\usepackage{caption}
\usepackage{subcaption}
\usepackage{xspace}
\usepackage{tcolorbox}
\usepackage{pifont}

\usepackage{enumitem}

\usepackage[colorlinks=true, linkcolor=black, urlcolor=blue, citecolor=black]{hyperref}

\renewcommand{\baselinestretch}{0.935}
\newcommand{\mxdotp}{\texttt{MXDOTP}\xspace}
\newcommand{\kone}{\textit{FP32}\xspace}
\newcommand{\ktwo}{\textit{FP8-to-FP32}\xspace}
\newcommand{\kmx}{\textit{MXFP8}\xspace}

\DeclareSIUnit[quantity-product = {}]\times{{$\times$}}

\newcommand{\cmark}{\ding{51}}  
\newcommand{\xmark}{\ding{55}}  

\newif\ifcameraready
\camerareadytrue 
\newcommand{\tbd}[1]{\textcolor{black}{#1}}
\newcommand{\changed}[1]{\protect\textcolor{black}{#1}}

\newif\ifarxiv
\arxivtrue

\ifarxiv
\usepackage{eso-pic}

\newcommand{\placetextbox}[3]{
  \setbox0=\hbox{#3}
  \AddToShipoutPictureFG*{
    \put(\LenToUnit{#1\paperwidth},\LenToUnit{#2\paperheight}){\vtop{{\null}\parbox[0pt][5pt][c]{\textwidth}{#3}}}%
  }%
}%
\fi

\def\BibTeX{{\rm B\kern-.05em{\sc i\kern-.025em b}\kern-.08em
    T\kern-.1667em\lower.7ex\hbox{E}\kern-.125emX}}
\begin{document}

\ifarxiv

\placetextbox{0.08}{0.05}{\textcolor{gray}{\footnotesize This paper has been accepted for publication by the 2025 IEEE 36th International Conference on Application-specific Systems, Architectures and Processors (ASAP). \copyright 2025 IEEE. Personal use of this material is permitted.  Permission from IEEE must be obtained for all other uses, in any current or future media, including reprinting/republishing this material for advertising or promotional purposes, creating new collective works, for resale or redistribution to servers or lists, or reuse of any copyrighted component of this work in other works.}}

\fi

\bstctlcite{IEEE:BSTcontrol}

\title{MXDOTP: A RISC-V ISA Extension for Enabling Microscaling (MX) Floating-Point Dot Products 
\ifcameraready
\\
\thanks{\changed{This work is supported in part by the NeuroSoC project funded under Horizon Europe Grant Agreement n° 101070634.}}
\fi
}

\author{
\IEEEauthorblockN{
Gamze İslamoğlu\IEEEauthorrefmark{1}, 
Luca Bertaccini\IEEEauthorrefmark{1}, 
Arpan Suravi Prasad\IEEEauthorrefmark{1},
Francesco Conti\IEEEauthorrefmark{2}, 
Angelo Garofalo\IEEEauthorrefmark{2}, 
Luca Benini\IEEEauthorrefmark{1}\IEEEauthorrefmark{2}
}
\IEEEauthorblockA{
\IEEEauthorrefmark{1}IIS, ETH Zurich, Switzerland; \IEEEauthorrefmark{2}DEI, University of Bologna, Italy \\
\{gislamoglu, lbertaccini, prasadar, lbenini\}@iis.ee.ethz.ch, 
\{f.conti, angelo.garofalo\}@unibo.it}
}

\maketitle

\newacronym[plural=CNNs, firstplural=Convolutional Neural Networks (CNNs)]{cnn}{CNN}{Convolutional Neural Network}
\newacronym[plural=DNNs, firstplural=Deep Neural Networks (DNNs)]{dnn}{DNN}{Deep Neural Network}
\newacronym[plural=LLMs, firstplural=Large Language Models (LLMs)]{llm}{LLM}{Large Language Model}
\newacronym[plural=SoCs, firstplural=System-on-Chips (SoCs)]{soc}{SoC}{System-on-Chip}
\newacronym[plural=ASICs, firstplural=Application-Specific Integrated Circuits (ASICs)]{asic}{ASIC}{Application-Specific Integrated Circuit}
\newacronym[plural=FMAs, firstplural=Fused Multiply-Add Units (FMAs)]{fma}{FMA}{Fused Multiply-Add}
\newacronym[plural=ViTs, firstplural=Vision Transformers (ViTs)]{vit}{ViT}{Vision Transformer}
\newacronym[plural=FPGAs, firstplural=Field-Programmable Gate Arrays (FPGAs)]{fpga}{FPGA}{Field-Programmable Gate Array}
\newacronym[plural=RNNs, firstplural=Recurrent Neural Networks (RNNs)]{rnn}{RNN}{Recurrent Neural Network}
\newacronym[plural=GPUs, firstplural=Graphics Processing Units (GPUs)]{gpu}{GPU}{Graphics Processing Unit}
\newacronym[plural=CPUs, firstplural=Central Processing Units (CPUs)]{cpu}{CPU}{Central Processing Unit}
\newacronym[plural=NPUs, firstplural=Neural Processing Units (NPUs)]{npu}{NPU}{Neural Processing Unit}
\newacronym[plural=FPUs, firstplural=Floating Point Units (FPUs)]{fpu}{FPU}{Floating Point Unit}
\newacronym[plural=ReLUs, firstplural=Rectified Linear Units (ReLUs)]{relu}{ReLU}{Rectified Linear Unit}
\newacronym{isa}{ISA}{Instruction Set Architecture}
\newacronym{simd}{SIMD}{Single Instruction Multiple Data}
\newacronym{rv}{RISC-V}{Reduced Instruction Set Computing - V}
\newacronym[plural=TPUs, firstplural=Tensor Processing Units (TPUs)]{tpu}{TPU}{Tensor Processing Unit}
\newacronym{ram}{RAM}{Random Access Memory}
\newacronym{lut}{LUT}{Lookup Table}
\newacronym{nlp}{NLP}{Natural Language Processing}
\newacronym{aiot}{AIoT}{Artificial Intelligence of Things}

\newacronym{dma}{DMA}{Direct Memory Access}

\newacronym[plural=CSRs, firstplural=Control Status Registers (CSRs)]{csr}{CSR}{Control Status Register}
\newacronym{ai}{AI}{Artificial Intelligence}
\newacronym{genai}{GenAI}{Generative AI}
\newacronym[plural=SSRs, firstplural=Stream Semantic Registers (SSRs)]{ssr}{SSR}{Stream Semantic Register}
\newacronym{frep}{FREP}{Floating-point Repetition}

\newacronym{mx}{MX}{Microscaling}
\newacronym{ocp}{OCP}{Open Compute Project}
\newacronym{bfp}{BFP}{block floating-point}

\newacronym{fp}{FP}{floating-point}
\newacronym{fprf}{FP RF}{floating-point register file}
\newacronym{rne}{RNE}{roundTiesToEven}
\newacronym{spm}{SPM}{Scratchpad Memory}
\newacronym{mac}{MAC}{Multiply-Accumulate}
\newacronym{mm}{MM}{matrix multiplication}

\begin{abstract}
Fast and energy-efficient low-bitwidth \gls{fp} arithmetic is essential for \gls{ai} systems.
\gls{mx} standardized formats have recently emerged as a promising alternative to baseline low-bitwidth \gls{fp} formats, offering improved accuracy with a block-wise shared exponent scale combined with per-element values.
However, efficiently executing the key linear algebra primitives for \gls{ai} applications on \gls{mx} formats requires specialized hardware support for the fundamental operators such as scaled dot product.
In this work, we propose \mxdotp, the first RISC-V ISA extension for \gls{mx} dot products, focusing on the 8-bit MXFP8 \gls{fp} format.
We extend the open-source Snitch RISC-V core with a dedicated MXFP8 dot product-accumulate unit, which fully consumes blocks of eight 8-bit operands packed into 64-bit inputs. 
To feed \mxdotp at full utilization with four operands per cycle, including block scales, we exploit Snitch's Stream Semantic Registers (SSRs), achieving up to \tbd{\qty[detect-all=true]{80}{\percent}} utilization with minimal impact on the Snitch core's architecture and no modification to the register file.
Implemented in \qty[detect-all=true]{12}{\nano\meter} FinFET, a cluster with eight \mxdotp-extended cores reaches up to \tbd{\qty[detect-all=true]{356}{\giga{FLOPS}\per\watt}} when computing MXFP8 matrix multiplications at \qty[detect-all=true]{0.8}{\volt}, \tbd{\qty[detect-all=true]{1}{\giga\hertz}}. Compared to a software baseline, where \gls{mx} dot products are computed by type casting FP8 inputs to FP32 for higher accumulation precision and applying explicit block scaling, the cluster achieves \tbd{\qty[detect-all=true,mode=text]{25}{\times}} speedup and \tbd{\qty[detect-all=true]{12.5}{\times}} better energy efficiency at a minimal \qty[detect-all=true]{5.1}{\percent} area increase.
\end{abstract}

\vspace{-0.5em}

\glsresetall

\section{Introduction}

Compact \gls{fp} arithmetic has gained significant attention for energy-efficient computing, particularly in machine learning workloads \cite{rouhani_pushing_2020, rouhani_microscaling_2023, lutz_fused_2024, bertaccini_minifloats_2024}. Deep learning models, like \glspl{llm}, demand massive computational and memory resources, making both compute efficiency and memory bandwidth key concerns. In low-bitwidth formats like FP8, increasing the exponent range improves dynamic range but reduces precision, while a larger mantissa improves precision at the cost of a smaller dynamic range. Block formats mitigate this hard tradeoff with a shared exponent, preserving both range and precision.

While \gls{bfp} formats have been widely studied \cite{rouhani_pushing_2020}, \gls{mx} formats offer greater flexibility with a two-level exponent representation, where groups of \gls{fp} elements share a wide exponent while each low-bitwidth element retains a narrower individual exponent to achieve finer-grained scaling. This reduces precision loss while retaining the advantages of low-bitwidth \gls{fp}.
\gls{mx} formats have demonstrated high accuracy in workloads such as \glspl{cnn}, \glspl{llm}, and \glspl{vit} \cite{rouhani_microscaling_2023}, leading to their adoption in a standard specification proposal developed by the \gls{ocp} \cite{rouhani_ocp_2023}.

Low-bitwidth formats significantly reduce memory and bandwidth needs by representing model parameters with fewer bits. However, fully realizing their efficiency gains requires native hardware support for scaled dot products, which dominate execution time in deep learning models. In large-scale data centers, reducing memory footprint and bandwidth is often prioritized, even at the cost of decompression overhead. In contrast, embedded \gls{ai} systems face strict power and compute constraints, where the overhead of format conversion can outweigh its benefits. This necessitates direct execution on \gls{mx} formats to maximize efficiency.

NVIDIA’s Blackwell Tensor Cores introduce micro-tensor scaling, incorporating community-defined \gls{mx} formats \cite{noauthor_nvidia_nodate}. Similarly, AMD’s Versal AI Edge Series Gen 2 employs MX6 and MX9 formats in its \gls{ai} engines \cite{noauthor_amd_nodate}. In the RISC-V ecosystem, Tenstorrent’s Tensix cores support \gls{bfp} formats \cite{noauthor_data_nodate}, but unlike \gls{mx}, they rely solely on shared exponents per block, lacking the finer-grained scaling of \gls{mx} (except for MXINT8). All these solutions are proprietary and tightly coupled to specialized tensor processing units. In contrast, our approach enables flexible and efficient execution of \gls{mx} computations on RISC-V cores without dedicated tensor accelerators.

Existing RISC-V architectures lack native support for efficient execution of mixed-precision \gls{mx} dot products due to the need for concurrent access to multiple types of data—scales, vector elements, and accumulators. Handling these dependencies in software requires frequent format conversions and explicit scaling operations, causing significant latency and energy overhead. To address these limitations, we propose \mxdotp, a dedicated \glsentryshort{isa} extension that natively integrates scaling within a dot product-accumulate operation. By eliminating redundant data movement and format conversions, \mxdotp enables high-throughput, energy-efficient execution of \gls{mx} matrix multiplications. The contributions of this paper are:


\begin{itemize} [leftmargin=12pt]
    \item We design a dot product-accumulate unit to support \gls{mx}FP8 (E5M2, E4M3) data formats with 8-bit scales and \changed{an FP32} accumulator to minimize overflows and precision losses.
    \item We introduce \mxdotp, a RISC-V \glsentryshort{isa} extension that fuses scaling and dot product into a single four-operand instruction. To sustain four operand fetches per cycle, we exploit \glspl{ssr} in the Snitch core~\cite{schuiki_stream_2021} to supply block scales without modifying the register file, which is typically limited to three read ports.
    \item We integrate \mxdotp into an 8-core Snitch cluster, reaching up to \tbd{\qty[detect-all=true]{80}{\percent}} utilization. Implemented in \qty[detect-all=true]{12}{\nano\meter} FinFET, it incurs only \qty[detect-all=true]{11}{\percent} core-level and \qty[detect-all=true]{5.1}{\percent} cluster-level area, and \qty[detect-all=true]{1.9}{\percent} power overhead when idle.
    \item We demonstrate a \tbd{\qty[detect-all=true,mode=text]{25}{\times}} speedup and \tbd{\qty[detect-all=true]{12.5}{\times}} improvement in energy efficiency over software-based \gls{mx} matrix multiplication on the baseline cluster.  
\end{itemize}

\begin{figure*}[t]
    \centering
    \vspace{-1.5em}
    \captionsetup{aboveskip=2pt}
    
    \includegraphics[width=0.99\linewidth]{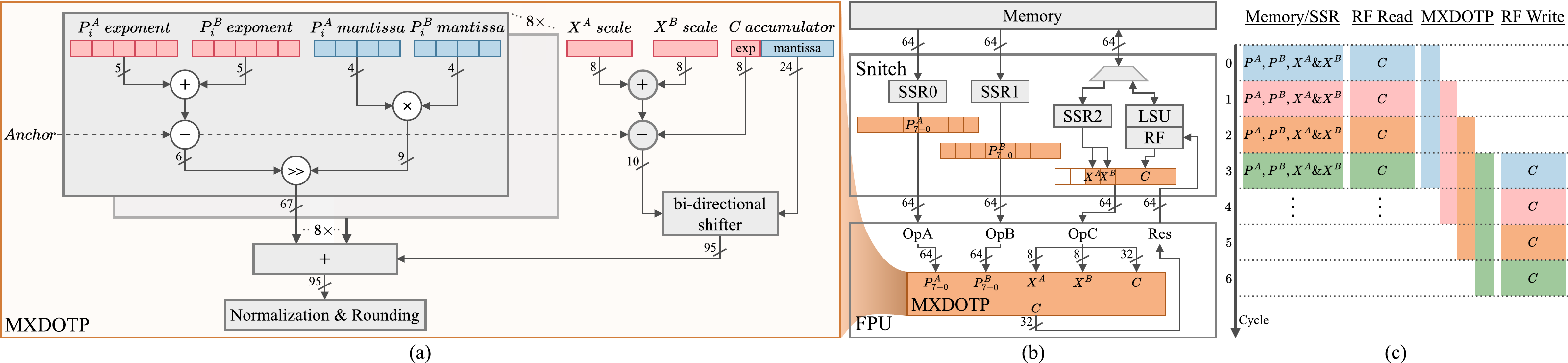}    
    \caption{(a) Simplified \mxdotp datapath (sign bits omitted). (b) Integration of \mxdotp into the Snitch core. (c) Execution of \mxdotp over multiple cycles with three pipeline stages. \glspl{ssr} enable efficient operand streaming, ensuring full utilization without stalls.}
\label{fig:mxdotp_snitch}
\vspace{-2em}
\end{figure*}

\section{Background}

\subsection{Microscaling (MX) Formats}
\gls{mx} formats, proposed by \gls{ocp} as a standard, are low-bitwidth data formats optimized for AI, co-developed by major industry players such as AMD, Arm, Intel, Meta, Microsoft, NVIDIA, and Qualcomm. These formats aim to enhance computational efficiency and memory footprint while maintaining accuracy in deep learning models. Unlike traditional per-tensor scaling or \gls{bfp}, which share a single exponent across large blocks, \gls{mx} offers finer-grained scaling by allowing individual elements or smaller blocks to retain independent scales. This reduces numerical errors while preserving memory efficiency. \gls{mx} serves as a drop-in replacement for FP32 in inference and enables sub-8-bit training for large generative models with negligible accuracy loss~\cite{rouhani_microscaling_2023}.

An \gls{mx}-compliant format consists of three components: \textit{a scale factor}, \textit{private elements}, and \textit{a block size}. The specification defines four concrete formats: MXFP8, MXFP6, MXFP4, and MXINT8. Each block of \( k \) elements shares a scale factor, while elements are encoded in FP8 (E5M2/E4M3), FP6 (E3M2/E2M3), FP4 (E2M1), or INT8. The scale is encoded using an E8M0 format with a fixed block size of 32. 

The \gls{mx} specification defines the dot product operation as:
\begin{equation}
    C = Dot (A,B) = X^{A}X^{B} \sum_{i=1}^k (P_i^A \times P_i^B) 
\end{equation}
where \( X^A \) and \( X^B \) are scale factors, and \( P_i^A \) and \( P_i^B \) are private elements of vectors \( A \) and \( B \). The internal precision of the dot product and the order of operations are implementation-defined. The general dot product for two \gls{mx}-compliant vectors is defined as:
\begin{equation}
    C = DotGeneral (A,B) = \sum_{j=1}^n Dot(A_j,B_j)
\end{equation}
where \( A \) and \( B \) consist of \( n \) \gls{mx} blocks of size \( k \), and result \( C \) \textit{should} be an FP32 number, as recommended by the \gls{mx} specification.

In this work, we focus on the MXFP8 general dot product to evaluate the efficiency of \gls{mx} formats in low-bitwidth arithmetic operations. Specifically, we utilize the two MXFP8 formats, E5M2 and E4M3,
to analyze their effects on power, performance, and area, \changed{as their FP8 counterparts are widely adopted in deep learning and supported by existing FPUs.}

    

\subsection{Snitch Cluster}
\label{sec:snitch_cluster}


The Snitch cluster is a high-performance, energy-efficient RISC-V multi-core architecture optimized for \gls{fp} workloads~\cite{zaruba_snitch_2021}. It features eight RISC-V RV32IMAFD cores, each equipped with a 64-bit \gls{fpu} supporting formats from FP64 to FP8. A \qty[detect-all=true]{128}{\kibi\byte} shared L1 \gls{spm} of 32 banks is interconnected via a single-cycle logarithmic interconnect, providing high-bandwidth, low-latency access. A dedicated DMA control core handles memory transfers between the \gls{spm} and external memory, reaching 512-bit peak bandwidth for data and instruction access through a 512-bit crossbar, while a separate 64-bit crossbar manages peripheral communication.

The cluster has two key \glsentryshort{isa} extensions.  
The \texttt{FREP} (Floating-point Repetition) extension~\cite{zaruba_snitch_2021} executes a predefined sequence of \gls{fp} instructions, removing the overhead of branching in tight arithmetic loops.
The \texttt{SSR} (Stream Semantic Register) extension~\cite{schuiki_stream_2021} eliminates explicit load/store instructions by enabling hardware-managed memory streams. Instead of issuing individual memory instructions, \glspl{ssr} autonomously fetch and store data with configurable affine access patterns, reducing address calculation overhead and instruction count. Each core has three \glspl{ssr}, programmable with a base address, stride, and loop bounds across up to four dimensions. After configuration, the \gls{ssr} logic manages memory transfers directly into the core registers, while the core focuses on computation. This improves utilization and overall efficiency.

\section{Architecture}

\subsection{MXDOTP Datapath and Integration into FPU}
\label{sec:datapath}

To efficiently implement the general dot product operation for the MXFP8 format, we adopt the principle of early accumulation introduced in \cite{lutz_fused_2024}. A simplified block diagram of the \mxdotp datapath is shown in \autoref{fig:mxdotp_snitch}a. To accommodate both FP8 formats, i.e. E5M2 and E4M3, we utilize an FP9 (E5M3) intermediate format. Instead of converting the sum of products to FP32 and adding it to the FP32 accumulator with a dedicated FP32 adder, early accumulation directly adds the shifted accumulator to the sum of products in fixed-point before the final conversion to FP32. The sum of multiplied elements and the accumulator is represented using a 95-bit fixed-point format with an anchor at 34, ensuring it can accommodate the full range of the sum of eight products along with the shifted accumulator, \changed{including sign and rounding bits}. Since the MX specification leaves internal precision as implementation-defined, we conservatively select the minimum bitwidth required to guarantee an exact result. We implement \gls{rne}, the most commonly used rounding mode in AI workloads, ensuring compliance with the specification, which leaves other modes optional.

The \mxdotp datapath is integrated into the \gls{fpu} of the Snitch cluster as an additional operation group\footnote{\changed{Available at: \url{https://github.com/pulp-platform/cvfpu/tree/feature/mxdotp}}}. Since the \gls{fpu} features a 64-bit data interface to support double-precision \gls{fp}, the \mxdotp datapath efficiently computes the dot product between vectors of eight elements with a throughput of one result per cycle, while its latency depends on the parametric number of pipeline stages.

\subsection{Instruction and Integration into Snitch}
\label{sec:snitch_integration}

We integrate the \gls{fpu} with \mxdotp into the Snitch core and extend the \glsentryshort{isa} with the \mxdotp instruction. The instruction operates on five inputs and produces an accumulated output, as detailed in \autoref{tab:mxdotp_operands}. Unlike a traditional three-operand \gls{fma} instruction, \mxdotp requires block scales as an additional input. However, the Snitch core and its \gls{fpu} are designed to handle up to three input operands, and the \gls{fprf} has three read ports and one write port, compliant with the RISC-V \glsentryshort{isa}. While adding another read port is a possible solution, it would introduce additional area overhead (around \qty[detect-all=true]{12}{\percent} to the \gls{fprf}). Instead, we leverage the \gls{ssr} extension of Snitch to stream scales without increasing area while also eliminating explicit load instructions for scales.


\begin{table}[t]
    \centering
    \vspace{0.7em}
    \captionsetup{aboveskip=3pt}
    
    \caption{Operands of the \mxdotp instruction.}
    \label{tab:mxdotp_operands}
    \resizebox{0.38\textwidth}{!}{
    \begin{tabular}{clcc}
        \toprule
        \textbf{Symbol} & \textbf{Operand} & \textbf{Role} & \textbf{Data Type} \\
        \midrule
        \( P^A \) / \( P^B \) & Elements of \( A \) / \( B \) & Input  & 8 × FP8 \\
        \( X^A \) / \( X^B \)& Scale factor of \( A \) / \( B \)& Input  & INT8 \\
        \( C \) & Accumulator & Input/Output & FP32 \\
        \bottomrule
    \end{tabular}
    }

\vspace{-0.6em}

\end{table}

\begin{table}[t]
    \centering
    \captionsetup{aboveskip=3pt}
    
    \caption{Encoding format of the \mxdotp{} instruction.}
    \label{tab:mxdotp_encoding}
    \resizebox{0.5\textwidth}{!}{
    \begin{tabular}{lccccccc}
        \toprule
        \textbf{Bits} & 31-27 & 26-25 & 24-20 & 19-15 & 14-12 & 11-7 & 6-0 \\ 
        \midrule
        \textbf{Field} & \texttt{rs3} & \texttt{sl} & \texttt{rs2} & \texttt{rs1} & \texttt{} & \texttt{rd} & \texttt{opcode} \\  
        \texttt{\textbf{\mxdotp}} & \( X^A \& X^B \)  & \(Select \) & \( P^B \) & \( P^A \) &  & \( C \) & 1110111 \\
        \bottomrule
    \end{tabular}
    }

\vspace{-2em}
\end{table}

\autoref{fig:mxdotp_snitch}b and c illustrate how operands are provided to the \gls{fpu} and the \mxdotp datapath. To efficiently manage operand fetching while adhering to register file constraints, we utilize three \glspl{ssr}. Two \glspl{ssr} stream the \( A \) and \( B \) vectors directly into the \gls{fpu} via its two 64-bit input interfaces, and the third \gls{ssr} streams both scale factors for \( A \) and \( B \). Since the scales require an additional operand beyond the standard three-operand FPU design, we merge the block scales with the FP32 accumulator and transfer them via the third 64-bit input interface of the \gls{fpu}. The final FP32 accumulated result is then written back to the \gls{fprf}. The only hardware modification required in the Snitch core for \mxdotp is the merging of scales with the accumulator.

A detailed representation of the encoding is shown in \autoref{tab:mxdotp_encoding} for the instruction \texttt{mxdotp rd, rs1, rs2, rs3, sl}. Since the register file only has three ports, this instruction requires at least one operand to be sourced via an \gls{ssr}. Bits 26-25 select the appropriate scales from the register \texttt{rs3}, as each 64-bit register stores four sets of scales. To distinguish between E5M2 and E4M3 FP8 formats, we introduce a dedicated \gls{csr}, which allows configuring the format prior to computation.

\definecolor{stringcolor}{HTML}{D1495B}   
\definecolor{mxdotpcolor}{HTML}{AD5A1C}   
\definecolor{convcolor}{HTML}{457B9D}     
\definecolor{ssrcolor}{HTML}{2E7D32}      


\lstdefinestyle{CStyle}{
    language=C,
    basicstyle=\ttfamily\fontseries{mb}\selectfont\tiny,  
    numbers=none,
    breaklines=true,
    tabsize=2,
    keywordstyle=\color{black},  
    morekeywords={mxdotp},      
    commentstyle=\color{gray},  
    showstringspaces=false,
    morekeywords={uint8_t},      
    stringstyle=\color{stringcolor},  
    moredelim=[is][\color{mxdotpcolor}\bfseries]{@}{@}, 
    moredelim=[is][\color{convcolor}]{!}{!},  
    moredelim=[is][\color{ssrcolor}]{&}{&},    
    moredelim=[is][\color{black}]{?}{?}    
}

\begin{figure*}[t]
  \centering
\vspace{-1.5em}
  \captionsetup{aboveskip=3pt}
    
  \begin{minipage}{0.31\textwidth}
      \lstset{style=CStyle}
      \begin{tcolorbox}[
          colframe=stringcolor!50, colback=white!98, 
          sharp corners=southwest, boxrule=0.7pt, width=\textwidth,
          boxsep=0.3pt, left=2pt, right=0pt, top=0pt, bottom=-6pt
      ]
      \caption*{\scriptsize \kone} 
      \vspace{-10pt} 
      \begin{lstlisting}
float *A, *B, *C; // FP32 input/output matrices
// Configure SSRs to stream A (ft0), B (ft1)
for (m = 0; m < M; m++):
  for (n = 0; n < N; n+=unroll): // unroll=8
    // Initialize SIMD vector with zeros
    // 8x (c0-c7)
    "vfcpka.s.s %[c0], %[zero], %[zero]"
    // Configure FREP hardware loop
    "frep.o %[n_frep], %[unroll], 0, 0"
    // Perform multiply-accumulate, 8x (c0-c7)
    "vfmac.s %[c0], ft0, ft1"
      \end{lstlisting}
      \end{tcolorbox}
  \end{minipage}
  \hfill
  \begin{minipage}{0.35\textwidth}
      \lstset{style=CStyle}
      \lstset{basicstyle=\ttfamily\fontsize{4.6}{5}\selectfont}
      \begin{tcolorbox}[
          colframe=convcolor!50, colback=white!98, 
          sharp corners=southwest, boxrule=0.7pt, width=\textwidth,
          boxsep=0.3pt, left=2pt, right=0pt, top=0pt, bottom=-6pt
      ]
      \caption*{\scriptsize \ktwo} 
      \vspace{-10pt} 
      \begin{lstlisting}
char *A, *B; float *C; // FP8 input, FP32 output matrices
uint8_t *Sa, *Sb;      // Scales
for (m = 0; m < M; m++):
  for (n = 0; n < N; n+=unroll): // unroll=8
    // Initialize accumulator fa1, fa2, ft8-ft11 (6 instrs)
    "lbu t1, 0(%[Sa])"      // Load Scale A
    "lbu t2, 0(%[Sb])"      // Load Scale B
    "add t3, t1, t2"        // Add scales
    "addi t3, t3, -127"     // Subtract bias
    "flh ft0, 0(%[A])"      // Load 2 FP8, 4x (ft0-ft3)
    "vfcvt.s.b ft0, ft0"    // Convert to FP32 (A), 4x
    "flh ft4, 0(%[B])"      // Load 2 FP8, 4x (ft4-ft7)
    "vfcvt.s.b ft4, ft4"    // Convert to FP32 (B), 4x
    "vfmac.s ft8, ft0, ft1" // 4x (ft8-ft11)
    "slli t3, t3, 23" // Shift scale to exponent
    "fmv.w.x ft0, t3" // Move scale (Sa + Sb) into fp reg             
    // Reduce result to fa2 (4 instrs)
    "fmadd.s fa1, fa2, ft0, fa1" // Scale result
      \end{lstlisting}
      \end{tcolorbox}
  \end{minipage}
  \hfill
  \begin{minipage}{0.31\textwidth}
      \lstset{style=CStyle}
      \begin{tcolorbox}[
          colframe=mxdotpcolor!50, colback=white!98, 
          sharp corners=southwest, boxrule=0.7pt, width=\textwidth,
          boxsep=0.3pt, left=2pt, right=0pt, top=0pt, bottom=-6pt
      ]
      \caption*{\scriptsize \kmx \texttt{(\mxdotp)}} 
      \vspace{-10pt} 
      \begin{lstlisting}
char *A, *B; // FP8 input matrices
float *C;    // FP32 output matrix
uint8_t *S;  // Scales
// Reshape scales (Sa and Sb to S) for SSR streaming
// Config SSRs to stream A (ft0), B (ft1), S (ft2)
for (m = 0; m < M; m++):
  for (n = 0; n < N; n+=unroll): // unroll=8
    // Initialize SIMD vector with zeros
    // 8x (c0-c7)
    "vfcpka.s.s %[c0], %[zero], %[zero]"
    // Configure FREP hardware loop
    "frep.o %[n_frep], %[unroll], 0, 0"
    // Perform MXDOTP instruction, 8x (c0-c7)
    @"mxdotp %[c0], ft0, ft1, ft2, 0"@
      \end{lstlisting}
      \end{tcolorbox}
  \end{minipage}

  \caption{Comparison of \kone, \ktwo, and \kmx \changed{\glsentryshort{mm}} kernels. $M$ and $N$ represent the output matrix’s row and column dimensions, respectively.}
  \label{fig:kernel_code}

\vspace{-1.4em}
\end{figure*}

\section{Results}
\label{sec:results}

\subsection{Implementation Setup and Area Analysis}
We synthesize, place, and route the \mxdotp-extended Snitch cluster in \textsc{GlobalFoundries'} \qty[detect-all=true]{12}{\nano\meter} FinFET technology with a target frequency of \qty[detect-all=true]{0.95}{\giga\hertz} under worst-case conditions (SS/\qty[detect-all=true]{0.72}{\volt}/\qty[detect-all=true]{125}{\celsius}) using Synopsys Fusion Compiler. To sustain frequency, the \mxdotp unit is implemented with three levels of pipeline registers. The cluster achieved \qty[detect-all=true]{1.09}{\giga\hertz} under typical conditions (TT/\qty[detect-all=true]{0.8}{\volt}/\qty[detect-all=true]{25}{\celsius}) without introducing any new critical paths. We use Synopsys’ PrimeTime to estimate power consumption under typical conditions, with switching activities extracted from a post-layout gate-level simulation at \tbd{\qty[detect-all=true]{1}{\giga\hertz}}. We average power consumption over ten samples extracted from DeiT-Tiny~\cite{touvron_training_2021}, quantized to MXFP8 using Microsoft's MX PyTorch Emulation Library\footnote{\url{https://github.com/microsoft/microxcaling}}. 

\begin{figure}[t] 
    \raggedright
    \captionsetup{aboveskip=2pt}
    \vspace{-0.5em}
    \includegraphics[width=0.96\linewidth]{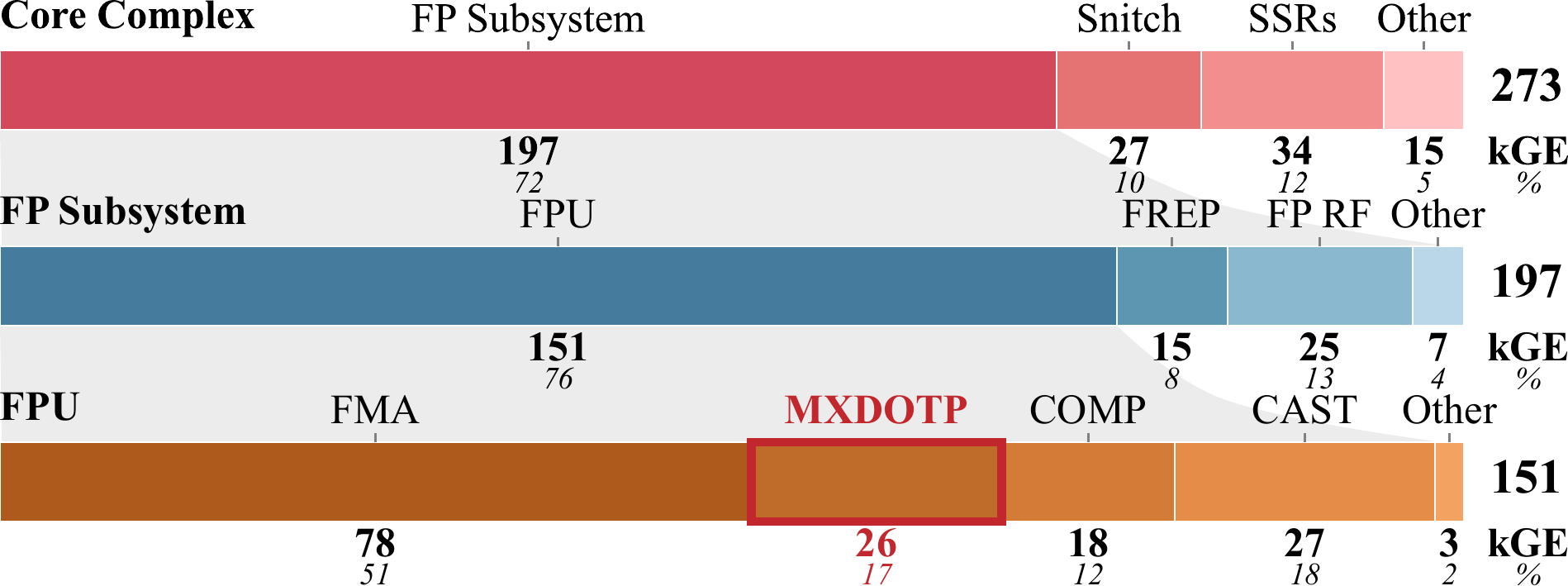} 
    \caption{Area breakdown of the \mxdotp-extended Snitch core.}
    \label{fig:area_breakdown}

\vspace{-1.2em}
\end{figure}

The total area of the cluster with \mxdotp-extended cores is \qty[detect-all=true]{4.89}{\mega{GE}}, representing a minimal \qty[detect-all=true]{5.1}{\percent} increase over the baseline Snitch cluster. The Snitch core complex consists of the core itself, its private instruction cache, \glspl{ssr}, and the \gls{fp} subsystem, which includes the \gls{fpu}, \glsentryshort{frep} logic, and \gls{fprf}. The area breakdown of these components is shown in \autoref{fig:area_breakdown}. \mxdotp accounts for \qty[detect-all=true]{17}{\percent} of the \gls{fpu} area and contributes \qty[detect-all=true]{9.5}{\percent} to the core complex. Moreover, it adds only \qty[detect-all=true]{1.9}{\percent} power overhead to the cluster when idle.

\subsection{Software Benchmark Setup}
To evaluate performance and energy efficiency, we compare three \changed{\gls{mm}} kernels: 
(1) \kone, a baseline kernel using standard 32-bit \gls{fp} multiply-accumulate (MAC) operations, 
(2) \ktwo, a software-based MX implementation that converts FP8 inputs to FP32, performs MAC operations, and applies block scale multiplications post-accumulation, and 
(3) \kmx, leveraging the proposed \mxdotp \glsentryshort{isa} extension to directly compute MX dot products in hardware, the block size remains configurable in software.

As shown in \autoref{fig:kernel_code}, the \kone kernel operates on FP32 inputs and accumulators using a 2-way SIMD MAC, while the \ktwo kernel introduces significant overhead due to FP8 to FP32 format conversions and explicit scale operations. In contrast, the \kmx kernel eliminates these overheads by executing the entire scaled dot product in a single instruction, efficiently integrating the scaling within the hardware pipeline.

\begin{figure}[t]
    \raggedright
    \captionsetup{aboveskip=0pt}
    
    \includegraphics[width=0.97\linewidth]{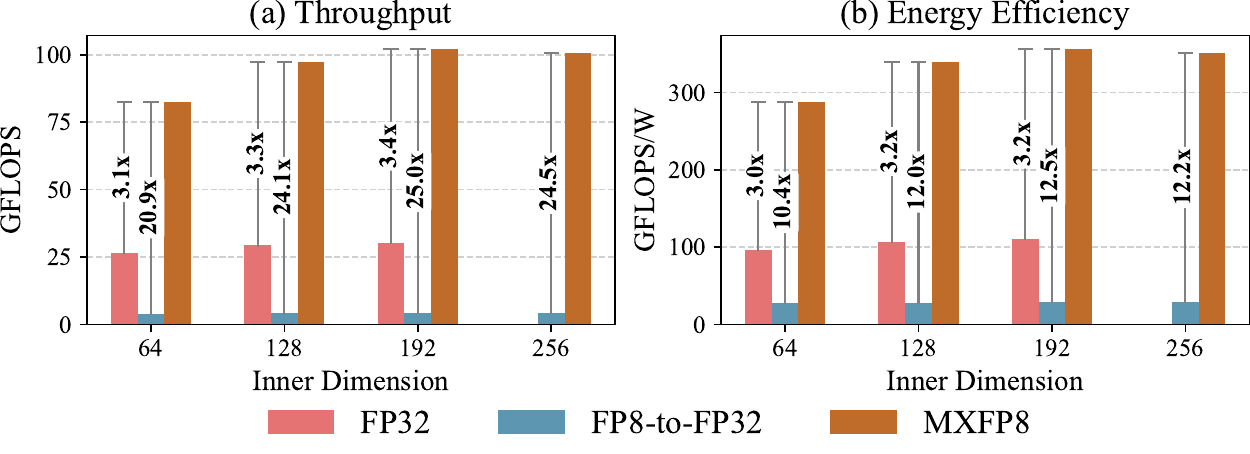} 
    \caption{(a) Throughput and (b) Energy Efficiency of \kone, \ktwo, and \kmx \changed{\glsentryshort{mm}} kernels for varying inner dimensions, with rows and columns fixed at 64. \kone does not fit into L1 with inner dimension of 256.}
    \label{fig:perf_energy}

\vspace{-1.9em}
\end{figure}

\subsection{Performance and Energy Efficiency}
As shown in \autoref{fig:perf_energy}, \mxdotp achieves a speedup of \tbd{\qty[detect-all=true,mode=text]{3.1}{\times} to \qty[detect-all=true,mode=text]{3.4}{\times}} over \kone and \tbd{\qty[detect-all=true,mode=text]{20.9}{\times} to \qty[detect-all=true,mode=text]{25.0}{\times}} over the \ktwo software baseline; it also improves energy efficiency by \tbd{\qty[detect-all=true,mode=text]{3.0}{\times} to \qty[detect-all=true,mode=text]{3.2}{\times}} and \tbd{\qty[detect-all=true,mode=text]{10.4}{\times} to \qty[detect-all=true,mode=text]{12.5}{\times}}, respectively. The \kmx kernel achieves up to \tbd{\qty[detect-all=true]{102}{\giga{FLOPS}}} and \tbd{\qty[detect-all=true]{356}{\giga{FLOPS}\per\watt}}, reaching \tbd{\qty[detect-all=true]{79.7}{\percent}} of the ideal throughput, factoring in \gls{ssr} and \glsentryshort{frep} configuration and loop overheads, accumulator initializations, and stores for final results.

While MX reduces memory footprint even without a dedicated dot product unit, our results show that the full performance and energy benefits of MX cannot be realized without hardware support for the general dot product operator. Compared to \kone, \mxdotp achieves higher throughput by leveraging lower bitwidth, allowing more data to be fetched and processed per cycle. Although \ktwo operates at the same low bitwidth, \mxdotp improves efficiency by fusing scaling and format conversions into a single operation. Moreover, frequent FP8 to FP32 format conversions and scale operations in \ktwo introduce significant computational overhead, making it less energy-efficient than even the \kone baseline. These findings highlight that dedicated MX dot product hardware is essential to fully unlock the performance and energy efficiency benefits of MX formats in \gls{ai} workloads. Implemented as an \glsentryshort{isa} extension, we provide full software flexibility, enabling seamless mixed-precision execution.

\subsection{Comparison to State-of-the-Art}

\begin{table}[t]
    \centering
    \captionsetup{aboveskip=1pt}
    \vspace{-0.3em}
    \renewcommand{\arraystretch}{1} 
    \setlength{\tabcolsep}{1pt} 
    \caption{Comparison of FP8 dot product units (first four rows) and compute clusters (last two rows).}
    \label{tab:soa}
    
    \resizebox{0.5\textwidth}{!}{%
    \begin{threeparttable} 
    
    \begin{tabular}{l S[table-format=2.0] S[table-format=1.1] S[table-format=1.2] 
                    c c c c S[table-format=4.0]}
    \toprule
    \textbf{Design} & \textbf{Tech.} & \textbf{Voltage} & \textbf{Freq.} & \textbf{Area} & \textbf{Scale} & \textbf{Accum.} & \textbf{Throughput\tnote{\changed{*}}} & \textbf{Efficiency\tnote{\changed{*}}} \\
                    & \si{\nano\meter}  & \si{\volt}     & \si{\giga\hertz} & \si{\milli\meter\squared} & \textbf{Support} & \textbf{Format} & \si{\giga{FLOPS}} & \si{\giga{FLOPS}/\watt} \\
    \midrule
    \textbf{ExSdotp\tnote{†}} ~\cite{bertaccini_minifloats_2024}  & 12 & 0.8  & 1.26  & \num{5.13e-3}  & \xmark & FP16  & 20.2   & 1631  \\
    \textbf{Desrentes \textit{et al.}\tnote{‡}} ~\cite{desrentes_exact_2023} & 16 & —  & 1.25  & \num{9.81e-3}  & \xmark & FP32  & 80.0   & 11300  \\
    \textbf{Lutz \textit{et al.}} \cite{lutz_fused_2024} & 5  & —  & 3.6   & \num{6.74e-4}  & \cmark\ (1 × 7b)  & FP32  & 28.8   & —  \\
    \textbf{This work}                                         & 12 & 0.8  & 1.09  & \num{3.15e-3}  & \cmark\ (2 × 8b)  & FP32  & 17.4   & 2035  \\
    \midrule
    \textbf{MiniFloat-NN} \cite{bertaccini_minifloats_2024} & 12 & 0.8  & 1.26  & 0.52     & \xmark  & FP16  & 128    & 575   \\
    \textbf{This work}                                         & 12 & 0.8  & 1.00  & 0.59     & \cmark\ (2 × 8b)  & FP32  & 102    & 356   \\
    \bottomrule
    \end{tabular}
\begin{tablenotes}
\item[*] \changed{1 FLOP = 1 \gls{fp} multiplication or 1 \gls{fp} addition.}
\item[†] Reported for FP8 with FP16 accumulation, though the unit also supports FP16 with FP32 accumulation.
\item[‡] We report the combined dot product (synthesized at \qty[detect-all=true]{250}{\mega\hertz} assuming it can be pipelined with four stages) and FP32 compression (synthesized at \qty[detect-all=true]{1.25}{\giga\hertz}) units for E5M2 format.
\end{tablenotes}
    \end{threeparttable}
    }
    
    \vspace{-2.2em}
\end{table}




We compare our work against prior FP8 dot product units and compute clusters in \autoref{tab:soa}. We first analyze standalone dot product units, corresponding to the first four rows of the table. ExSdotp~\cite{bertaccini_minifloats_2024} performs FP8 dot products with FP16 accumulation, while Desrentes et al.~\cite{desrentes_exact_2023} support exact accumulation and FP32 conversion. However, both lack scaling support, whereas Lutz et al.~\cite{lutz_fused_2024} introduce a single 7-bit scale. Our design advances beyond these approaches by incorporating two independent 8-bit scales, fully enabling MX general dot product. With an energy efficiency of \qty[detect-all=true]{2}{\tera{FLOPS}/\watt}, it outperforms ExSdotp. The higher efficiency of \cite{desrentes_exact_2023} is primarily due to three factors: (1) our design includes additional scaling operations, which are not counted as OPs, (2) \cite{desrentes_exact_2023} reports synthesis estimates at different frequencies, while we provide post-layout area and power results, and (3) we report only E5M2 results for \cite{desrentes_exact_2023}, as their design does not support a combined unit for E5M2 and E4M3, unlike \mxdotp.

At the cluster level (last two rows), our implementation achieves competitive throughput and efficiency compared to MiniFloat-NN~\cite{bertaccini_minifloats_2024}, which integrates ExSdotp as an \glsentryshort{isa} extension. However, MiniFloat-NN does not include a dedicated mechanism for handling block scaling, requiring an additional software stage. In contrast, our approach natively integrates block scaling within the dot product operation while achieving the same frequency-normalized throughput as MiniFloat-NN. 


\section{Conclusion}
\label{sec:conclusion}  

We presented \mxdotp, a RISC-V \glsentryshort{isa} extension enabling efficient execution of \glsentrylong{mx} (\glsentryshort{mx}) dot products. By integrating a dedicated dot product-accumulate unit into the Snitch core and leveraging \glspl{ssr} to handle scaling factors, we enable hardware acceleration of MX-based matrix multiplications. Implemented in a \qty[detect-all=true]{12}{\nano\meter} FinFET technology, our design achieves \tbd{\qty[detect-all=true]{102}{\giga{FLOPS}}} and \tbd{\qty[detect-all=true]{356}{\giga{FLOPS}\per\watt}}, \tbd{\qty[detect-all=true]{25}{\times}} speedup and \tbd{\qty[detect-all=true]{12.5}{\times}} improvement in energy efficiency over a software-based \gls{mx} \changed{\glsentrylong{mm}} kernel. These results highlight the necessity of dedicated \gls{mx} \glsentryshort{isa} and micro-architectural support for aggressively block-quantized AI workloads.

\bibliographystyle{IEEEtran}
\bibliography{references}

\end{document}